\documentclass[aps,prl,showpacs,twocolumn]{revtex4}

\usepackage{amsmath,epsfig,amsfonts,graphicx,color,ulem}
\usepackage{graphicx}

\begin{document}
\title{Dissipative Solitary Waves in Granular Crystals}
\author{
R. Carretero-Gonz\'alez$^{1}$,
D. Khatri$^{2}$,
Mason A. Porter$^{3}$,
P. G. Kevrekidis$^{4}$, and
C. Daraio$^{2,}$\footnote{corresponding author}
}

\affiliation{
$^1$%
Department of Mathematics and Statistics,
San Diego State University, San Diego CA, 92182-7720, USA\\
$^2$Graduate Aeronautical Laboratories (GALCIT) and Department of Applied Physics,
California Institute of Technology, Pasadena, CA 91125, USA \\
$^3$Oxford Center for Industrial and Applied Mathematics,
Mathematical Institute, University of Oxford, OX1 3LB, UK \\
$^4$Department of Mathematics and Statistics,
University of Massachusetts, Amherst MA 01003-4515, USA
}

\begin{abstract}
We provide a quantitative characterization of dissipative effects in
one-dimensional granular crystals.  We use the propagation
of highly nonlinear solitary waves as a diagnostic tool and develop
optimization schemes that allow one to compute the relevant exponents
and prefactors of the dissipative terms in the equations of motion.
We thereby propose a quantitatively-accurate extension of the Hertzian
model that encompasses
dissipative effects via a discrete Laplacian of the velocities.  Experiments
and computations with steel, brass, and polytetrafluoroethylene reveal
a {\it {common}} dissipation exponent
with a material-dependent prefactor.
\end{abstract}



\maketitle

\noindent{{\sl \em Introduction.}}  {{Since the advent of the famous
Fermi-Pasta-Ulam model over fifty years ago, nonlinear oscillator chains have received a remarkable
amount of attention in a wide range of physical settings
\cite{fpu55-focus}. Areas of intense theoretical
and experimental interest over the last decade include
(but are not limited to)
DNA double-strand dynamics in biophysics \cite{peyrard},
coupled waveguide arrays
in nonlinear optics \cite{photon}, breathing oscillations
in micromechanical cantilever arrays \cite{sievers}, and
Bose-Einstein condensation in optical lattices in atomic
physics \cite{morsch1}. }}

Within this general theme of interplay between nonlinearity
and discreteness, one of the key subjects
has been the study of one-dimensional (1D) granular
materials, which consist of chains of interacting particles that start from point contact with each other and deform elastically when compressed.  In contrast to classically-studied disordered granular media, highly-packed granular
lattices have negligible frictional and rotational dynamics, in favor of axial stress propagation \cite{sen08,s2}.  The highly nonlinear
dynamic response of such ``crystals'' has been the subject of
considerable attention \cite{nesterenko1,sen08,coste97,nesterenko2,dar05,dar05b,dar06,hascoet00,hinch99,man01,hong02,hong05,job05,doney06,ver06,sok07,herb06}.
Additionally, granular crystals can be created from numerous
material types and sizes, which makes
their properties extremely tunable \cite{nesterenko1,nesterenko2,coste97}.
 This flexibility is valuable not only for
basic studies of the underlying physics but also in potential
applications such as shock \cite{ericshock} and energy absorbing
layers \cite{dar06,hong05,doney06,ver06}, sound focusing devices (tunable
acoustic lenses and delay lines), actuators \cite{dev08,patent},
sound absorption layers, and sound
scramblers \cite{dar05,dar05b,herb06}.

While the standard Hertzian force model has been used
extensively in most dynamical investigations and is now textbook
material \cite{nesterenko1,s3}, recent experimentally-motivated
investigations have illustrated the challenging need to include
dissipation effects
\cite{others,others2,ericshock,LindenbergNesterenkoPRL07}.
Dissipative terms related to friction \cite{wang08}, plasticity
\cite{pane05}, visco-elasticity \cite{morgado97}, and viscous drag
\cite{LindenbergNesterenkoPRL07,herb06}
have been proposed to model particle collisions \cite{sen08,s2,s3}.
However, none of these models captures both
qualitatively and quantitatively the decay and wave shape of
the highly nonlinear solitary waves observed experimentally.
It is this important experimental and
theoretical aspect of packed granular lattices that we aim to tackle
in this Letter through the combination of modeling, numerical and physical
experiments, and a detailed comparison thereof.  Based on the earlier
propositions of Refs.~\cite{ericshock,LindenbergNesterenkoPRL07,sen08,s2},
we illustrate the {prevalent} nature of dissipation in the form of a discrete Laplacian
in the velocities with {\it uniform} exponent and a material-dependent prefactor.
The broad interest of our findings results not only from their {general nature}
for granular crystals of different materials but also from
the significance of similar models in other fields,
such as 1D lattice turbulence \cite{michel}.
%

\begin{figure}[tbp]
\begin{center}
\includegraphics[width=3.85cm]{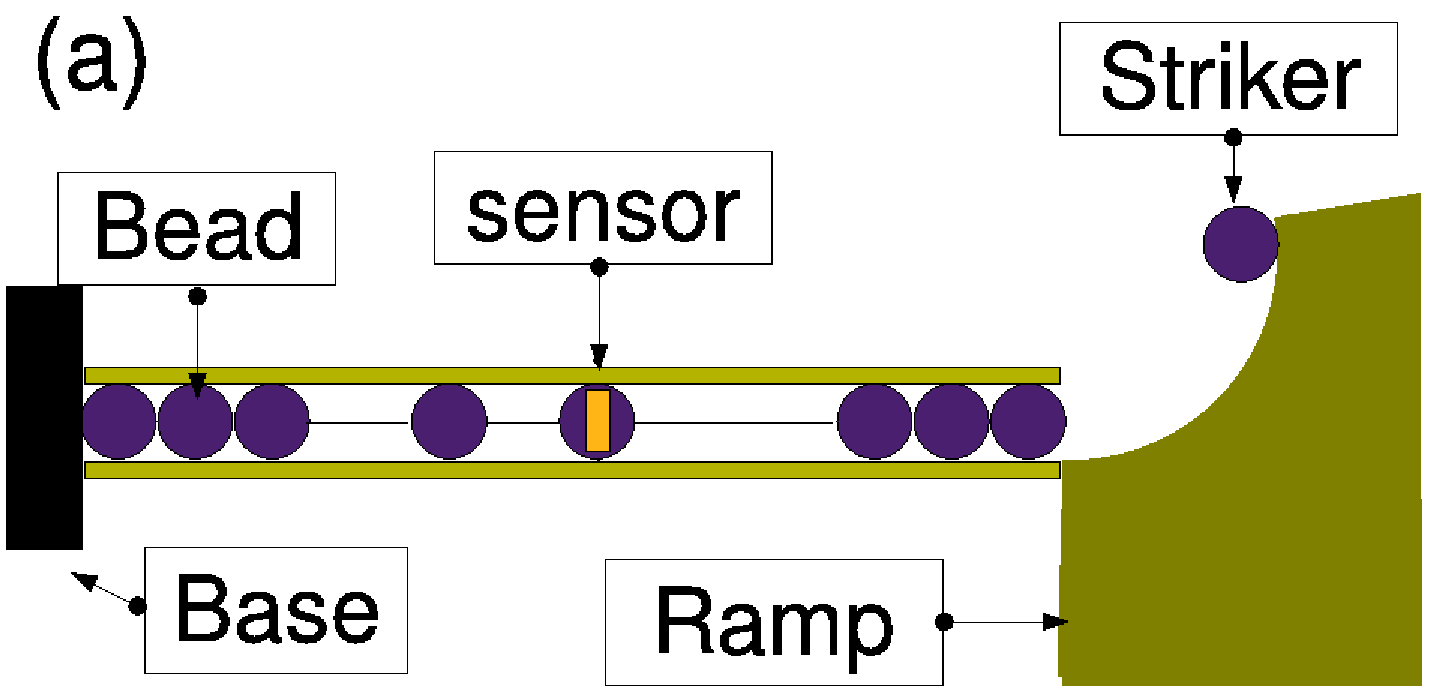}
~
\includegraphics[width=4.50cm]{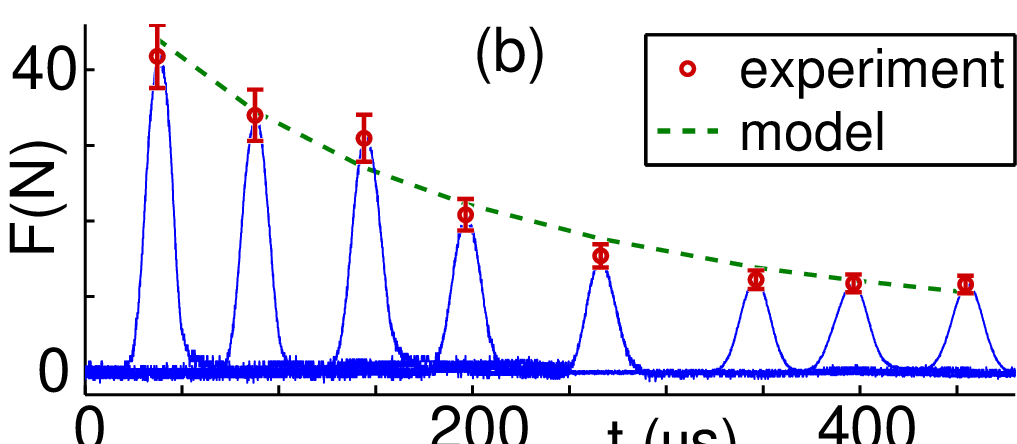}
\end{center}
\vspace{-0.4cm}
\caption{(Color online)
(a) Schematic diagram of the experimental setup.
(b) Solitary wave decay in a chain composed of 70 steel
particles impacted by a steel bead with $v_{{\rm imp}} = v_1$.
The (blue) solid curves correspond to the recordings for sensors placed
in particles 9, 16, 24, 31, 40, 50, 56, and 63.
}
\label{Hchain}
\end{figure}

\noindent{{\sl \em Experimental Setup.}}
We assembled a monodisperse chain of $N$ beads
(here we report results for $N=70$ but we performed experiments
for up to $N=188$ with similar results)
of different materials (see Table \ref{parameters})
with radius $R = 2.38$ mm in a horizontal setup (see Fig.~\ref{Hchain}a)
composed of four-garolite rod stand.
(To ensure
contact between the particles, the guide was tilted at 4 degrees.)
To directly visualize the waves, we embedded calibrated piezo sensors
($RC\sim 10^3 \mu s$, Piezo Systems Inc; see Fig.~1b of Ref.~\cite{dar05})
inside selected particles, as described in Refs.~\cite{nesterenko2,dar05,dar05b,dar06}.
We generated solitary waves by impacting the chain with a striker
(identical to the particles in the chain) launched along a ramp.
We calculated the impact velocities $v_{{\rm imp}}$ (in m/s) using
a high-speed camera at the end of the ramp:
$v_{1,2}=1.77$,
$v_{3,4}= 1.55$,
$v_{5,6}= 1.40$,
$v_{7,8}= 1.04$, and
$v_{9,10}= 0.79$.

%
%
%

\begin{table}
\centerline{
\begin{tabular}{|c|c|c|c|c|c|} \hline
Material & $m$ (g) & $E$ (GPa) & $\nu$ & $\alpha$ & $\gamma$ \\ \hline
Steel    & 0.45    & 193       & 0.30  & $1.81\pm 0.25$ & $-5.58\pm 1.30$ \\ \hline
PTFE     & 0.123   & 1.46      & 0.46  & $1.68\pm 0.16$ & $-1.56\pm 0.19$ \\ \hline
Brass    & 0.48    & 103       & 0.34  & $1.85\pm 0.13$ & $-6.84\pm 0.66$ \\ \hline
\end{tabular}}
\caption{Material properties (mass $m$, elastic modulus $E$,
and Poisson ratio $\nu$) for stainless steel \cite{metals,316},
PTFE \cite{dar05,dupont,carter95}, and brass \cite{brass}.
The last two columns present our best estimates, together with their
standard deviation, of the dissipation coefficients $(\alpha,\gamma)$.
}
\label{parameters}
\vspace{-0.4cm}
\end{table}


\noindent{{\sl \em Model.}}
We model a dissipative chain of $N$ spherical beads as a 1D lattice with
Hertzian interactions \cite{nesterenko1}:
\begin{align}
    \ddot{y}_n &= A\left(\delta_{n}^{3/2}-\delta_{n+1}^{3/2}\right)
    +\gamma s
   \left|\dot\delta_{n}-\dot\delta_{n+1}\right|^\alpha
\,, \label{motion}
\end{align}
where $s\equiv {\rm sgn}(\dot\delta_{n}-\dot\delta_{n+1})$,
$A\equiv{E\sqrt{2R}}/[{3m\left(1-\nu^2\right)}]$,
$n \in \{1,\dots,N\}$, $y_n$ is the deviation of the
$n$th bead from its equilibrium,
$\delta_n \equiv \mbox{max}\{y_{n-1} - y_{n},0\}$
for $n \in \{2,\dots,N\}$, $\delta_1 \equiv 0$,
$\delta_{N+1} \equiv \mbox{max}\{y_{N},0\}$,
$E$ is the Young's (elastic) modulus of the beads,
$\nu$ is their Poisson ratio, $m$ is their mass, and $R$ is their radius.
The particle $n = 0$ represents the striker.
Dissipation is incorporated by using a
phenomenological force, of prefactor $\gamma<0$,
between adjacent beads that depends on
their relative velocities (in particular, on the ``discrete Laplacian''
in the velocities), generalizing earlier models (with dissipation
exponent $\alpha = 1$
specifically) for dry granular matter \cite{LindenbergNesterenkoPRL07}.
%

%
In contrast to previous works that a priori assume
that $\alpha = 1$ (i.e., that model dissipation using a linear
dashpot) \cite{LindenbergNesterenkoPRL07}, we determine both
$\alpha$ and $\gamma$ by directly comparing experimental and
numerical results. The general coefficient $\alpha$
is thus a phenomenological parameter derived from the best fitting.
We introduce the absolute value and the sign parameter $s$ into
(\ref{motion}) to ensure that genuine dissipation is guaranteed
irrespective of the sign of the relative velocities between consecutive
beads. The units of $\gamma$ would depend on the value of $\alpha$
and, accordingly, are more properly investigated in dynamic models
that incorporate dissipation based on detailed measurements of
restitutive losses that cannot currently be achieved experimentally \cite{sen08}.
Importantly, the value we obtain for $\alpha$ differs decidedly
from the coefficients used in previous modeling attempts
\cite{LindenbergNesterenkoPRL07} (see the discussion below).
%

\begin{figure}[tbp]
\centering{
~~\includegraphics[width=4.1cm]{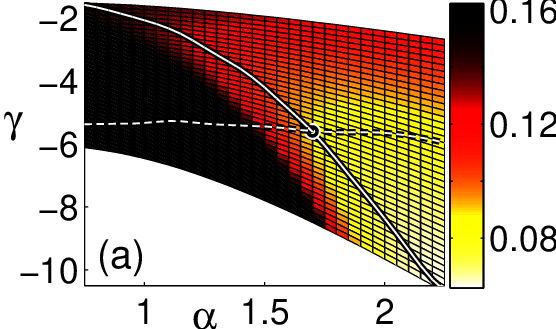}
~\includegraphics[width=4.1cm]{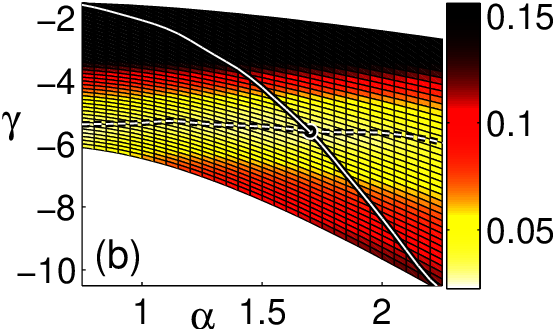}\\
\includegraphics[width=4.0cm]{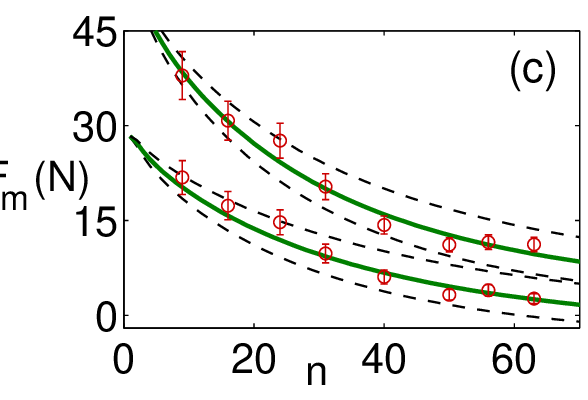}~~
\includegraphics[width=4.0cm]{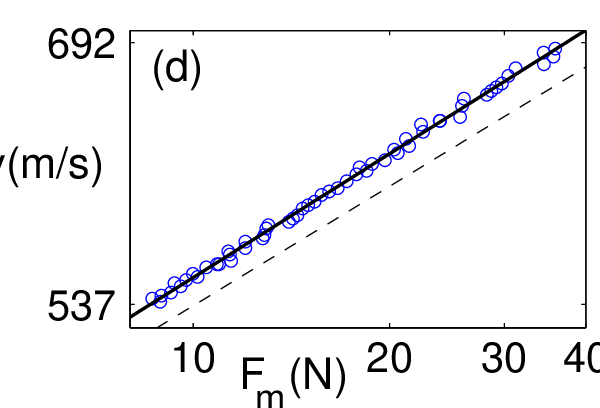}
\vspace{-0.2cm}
\vspace{-0.2cm}
}
\caption{(Color online)
Optimization of the dissipation coefficients $(\alpha,\gamma)$
for a steel chain.
(a) Difference $D(\alpha,\gamma)$, as defined in Eq.~(\ref{diff1}),
between the force maxima recorded in the experiment and our model.
(b) Difference $\Delta_n(\alpha,\gamma)$, as
defined in Eq.~(\ref{diff2}), in wave forms between the
experiment and our model for sensor $n=56$.
The solid and dashed curves
correspond to the minima obtained from panels (a) and (b), respectively.
(c) Maximum force $F_m(n)$ for experiments with $v_{{\rm imp}}=v_{3}$
(top curves) and $v_{{\rm imp}}=v_{8}$
(bottom curves, displaced by 5 units for clarity).
The (red) circles
correspond to the experiment,
and the (green) thick curves give the numerical best fit
with $(\alpha,\gamma)=(1.81\pm 0.25,-5.58\pm 1.30)$.
The dashed curves correspond to the extreme cases using
the standard deviation found in the optimal parameters.
(d) Velocity of traveling front versus the maximum force (in a log-log
plot). The solid curve represents the best linear fit, which gives
{$v \propto F_m^{0.17}$; we also show a dashed line with slope $1/6$.}
}
\label{fig2}
\end{figure}


\vspace{.1 in}

\noindent{{\sl \em Determining the dissipation coefficients.}}  We now determine the ``optimal'' dissipation coefficients $(\alpha,\gamma)$
from the experimental data for different materials and different
configurations. The experimental data consists of the time series
of the force through each sensor. We optimize the pair
$(\alpha,\gamma)$ by minimizing the following two
differences between numerics and each particular experiment:
\begin{align}
    D(\alpha,\gamma) &\!\!=\!\! \frac{1}{N} \sum_{n=1}^{N}
\frac{\left|F_m^{\rm exp}(n) - F_m^{\rm num}(n)\right|}{\bar{F}_m^{\rm exp}},
\label{diff1}
\\[1.0ex]
\Delta_n(\alpha,\gamma) &\!\!=\!\! \frac{1}{T}
 \int_{t_i}^{t_f}\!
\frac{\left|F^{\rm exp}(t;n) - F^{\rm num}(t;n)\right|}{\bar{F}^{\rm exp}(n)}\, dt\,, \label{diff2}
\end{align}
%
where $\bar{F}_m^{\rm exp}\equiv (1/N)\sum_{n=1}^{N}  F_m^{\rm exp}(n)$,
$\bar{F}^{\rm exp}(n)=\int_{t_i}^{t_f}F^{\rm exp}(t;n)dt$,
$F(t;n)$ is the time series data of the force through the $n$th sensor (see Fig.~\ref{Hchain}b),
and $F_m(n) = \max_t\{F(t;n)\}$ is the maximum force recorded by the $n$th sensor over the recording time span $[t_i,t_f=t_i+T]$, where $T$ is typically about 100 $\mu$s.
The superscripts `exp' and `num' denote, respectively, the experimental and numerical data.  The function $D(\alpha,\gamma)$ measures the ``distance'' between the numerics and the experiment using the {\it maxima} of the forces through all sensors of the experiment.  The function $\Delta_n(\alpha,\gamma)$ measures the difference between experimental and numerical
pulse shapes that go through the $n$th sensor.  In order to avoid biasing $\Delta_n(\alpha,\gamma)$ with the difference in force magnitude [which is already taken into
account when optimizing $D(\alpha,\gamma)$], we rescale the experimental data so that the numerical and experimental maxima match before we compare wave forms.  That is, $F^{\rm exp}(t;n) \rightarrow F^{\rm exp}(t;n)\times F_m^{\rm num}(n)/F_m^{\rm exp}(n)$.  Panels (a) and (b) in Fig.~\ref{fig2} depict, respectively, the differences $D(\alpha,\gamma)$ and $\Delta_n(\alpha,\gamma)$ in
a particular $(\alpha,\gamma)$ range for a steel chain using a sensor placed towards the end of the chain.  As can be observed from these panels, the optimization of the force maxima $F_m$ [panel (a)] and the force pulse shape [panel (b)] are not sufficient on their own to determine the dissipation parameters.  However, it is meaningful (and always well-defined) to optimize force maxima {\it and} pulse shape together by taking the intersection between the minima of
each case (see the point at the intersection of the solid
and dashed curves). For experiment $j$ (with impact velocity $v_j$), we average
the parameter pair $(\alpha_j,\gamma_j)$ over four sensors located throughout the
bead chain. Finally, we average $(\alpha,\gamma)=\frac{1}{N_e}\sum_{j=1}^{N_e}
(\alpha_j,\gamma_j)$ over the $N_e=10$ different experiments to obtain
the optimal dissipation parameters $(\alpha,\gamma)$ and compute
the standard deviation for the $N_e$ experiments.

We summarize our results, for three different set of experiments---using steel, teflon (polytetrafluoroethylene; PTFE), and brass beads---in the last two columns of Table~\ref{parameters}.  In order to validate the results of the above optimization procedure a posteriori, we take the optimal dissipation parameters for the steel bead chain and compare the maximal forces obtained numerically with the experiments in panel (c) of Fig.~\ref{fig2}. In the panel, we show two typical examples (for impact velocities $v_3$ and $v_8$) and also plot the curves incorporating the standard deviation measured in
our analysis. As can be clearly observed,
all experimental data points fall well within the predicted region.  To further validate our results, we compared the dependence of the pulse velocity $v$ against the maximal force $F_m$ in panel (d) of Fig.~\ref{fig2} [this panels shows a typical example; we obtained similar results for the other configurations (results not shown here)].
The obtained exponent is extremely close to the theoretical value of $1/6$ (shown by the dashed line) \cite{nesterenko1}.

\begin{figure}[tbp]
\centering{
\includegraphics[width=4.05cm,height=2.4cm]{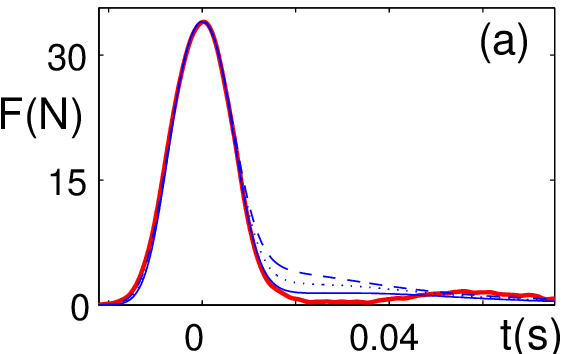}~
\includegraphics[width=4.05cm,height=2.4cm]{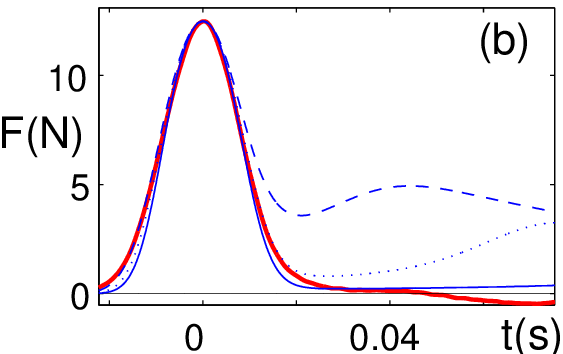}
\vspace{-0.2cm}
}
\caption{(Color online)
Force versus time for the steel chain with $v_{{\rm imp}}=v_2$ through sensors
at positions (a) $n=16$ and (b) $n=56$.  The (red) thick solid curve depicts
the (smoothed; see text) experimental series, and the thin (blue) dashed,
dotted, and solid curves respectively show the numerics with
$(\alpha,\gamma)=(1,-5.5)$, $(1.4,-6)$, and $(1.81,-5.58)$.
The last case corresponds to the best fit (see text) for the
dissipation parameters for the chains of steel beads.
}
\label{fig3}
\end{figure}

In order to gain a deeper understanding of the role of the
dissipation exponent $\alpha$, we depict in Fig.~\ref{fig3}
the pulse shape for two sensors in the steel chain
(one near the beginning of the chain and the other one near the end).
We depict the
experimental pulse (smoothed by nearest-neighbor averaging)
with the (red) solid curve. The thin (blue) curves,
depict three numerical runs using three different pairs
($\alpha$,$\gamma$) along the minimum curve [shown by a
solid curve in panel (a) of Fig.~\ref{fig2}].

It is interesting to note that for all materials tested, higher
impact velocities correspond to a faster initial
decay
as compared to the latter part of the chain {(probably related
to the initiation of plasticity at the contact)}.  Also, by comparing
the wave decay in chains composed of steel and teflon (or brass)
beads, a faster and more pronounced energy loss is evident
for the softer beads. To understand physically this
dissipation, one should explore a more detailed analysis of the
contact plasticity, inelastic restitution, and hydrodynamic drag.
We stress here that our granular crystals have a closely packed
particle arrangement and limited or null rotational dynamics
and particles' displacements (small frictional response).

Note that the optimal dissipation exponent $\alpha$ for the
three material types considered is consonant with a value
close to $\alpha=1.75$. This indicates the {\it {prevalence}}
of the phenomenological damping introduced in Eq.~(\ref{motion}),
which is one of the principal findings of this Letter.
It is important to point out the disparity of this optimal
exponent from earlier investigations,
which focused on the (linear dashpot) case of $\alpha=1$
\cite{ericshock,LindenbergNesterenkoPRL07,michel}.
On the other hand, naturally, the dissipation prefactor $\gamma$ does
depend on the material. For steel and brass, which
have similar material properties, $\gamma$ is also similar
(steel has $\gamma = -5.58$ and brass has $\gamma = -6.84$).
However, for teflon, as can be anticipated from the much lower
elastic modulus $E$, the prefactor $\gamma$ is significantly
smaller ($\gamma = -1.56$).

\begin{figure}[tbp]
\centering{
  \includegraphics[width=4.10cm,height=2.6cm]{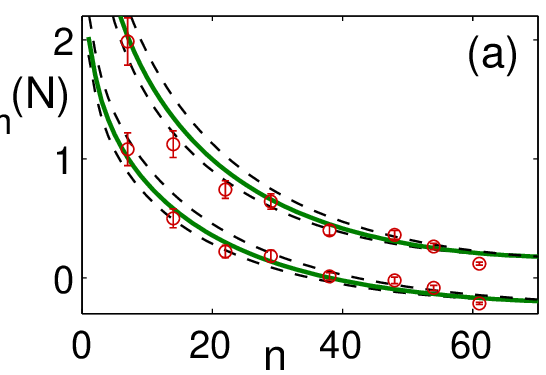}~
  \includegraphics[width=4.10cm,height=2.6cm]{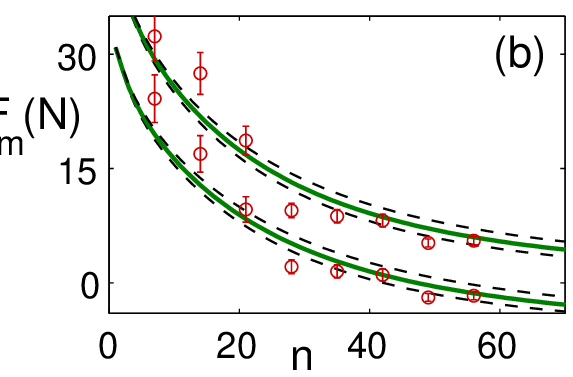}\\[1.0ex]
~\includegraphics[width=4.15cm,height=2.3cm]{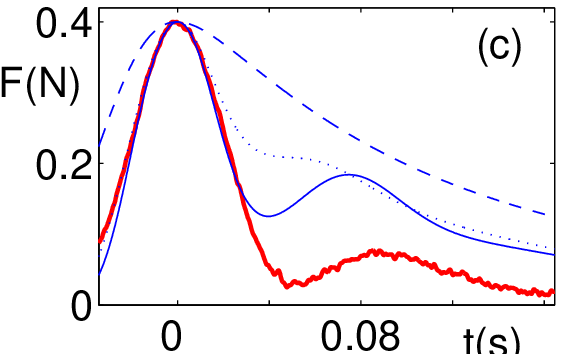}~
  \includegraphics[width=4.05cm,height=2.3cm]{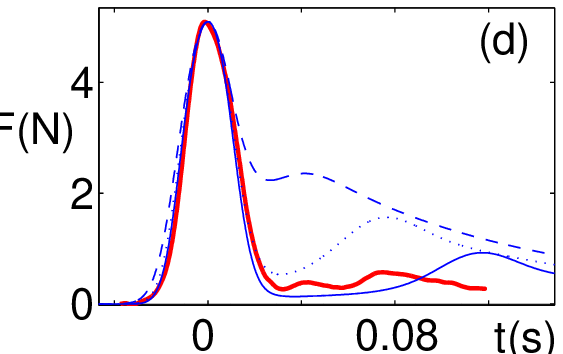}
\vspace{-0.2cm}
}
\caption{ (Color online)
Results for the teflon (left column) and brass (right column) experiments.
The top panels depict the same information as panel (c) in Fig.~\ref{fig2}
for experiments with impact velocities (a) $v_3$ (top curves) and $v_8$
(bottom curves, displaced by 0.3 units for clarity); and (b) velocities
$v_3$ (top curves) and $v_6$ (bottom curves, displaced by 7 units for clarity).
The best fit for the dissipation parameters for teflon and brass
are, respectively, $(\alpha,\gamma)=(1.68\pm 0.16,-1.56\pm 0.19)$ and
$(\alpha,\gamma)=(1.85\pm 0.13,-6.84\pm 0.66)$.  The bottom panels show
the same information as in Fig.~\ref{fig3}. In panel (c), we depict the
force versus time through the sensor at $n=38$ with
$(\alpha,\gamma)=(1,-1.56)$, $(1.4,-1.56)$, and $(1.68,-1.56)$.
In panel (d), we show the same information for the sensor at
$n=14$ with $(\alpha,\gamma)=(1,-5.5)$, $(1.4,-6)$, and $(1.85,-6.84)$.
}
\label{fig4}
\end{figure}

%
We show typical examples of the results for teflon (left column)
and brass (right column) in Fig.~\ref{fig4}.  The top panels
depict the maximal force through the chain using the optimal
dissipation parameters.
Note in the pulse shape results (bottom panels)
for teflon and brass that low dissipation
exponents $\alpha$ tend to overestimate the size of the
secondary pulse hump.
Another relevant
observation, in connection with its much smaller dissipation
prefactor $\gamma$, is that chains of teflon beads may offer
the first unambiguous observation of the {\it secondary pulses}
(see Fig.~\ref{fig4}c) argued to arise for weaker
dissipation in Ref.~\cite{LindenbergNesterenkoPRL07}.


\vspace{.1 in}

\noindent{{\sl \em Conclusions.}}
In this Letter, we have offered for the first time a
quantitative and systematic modeling attempt at the role of dissipation
in granular crystals. Through detailed comparison of numerical
simulations and experiments in a variety of materials
(steel, teflon, and brass), we have demonstrated a {generic} functional
form of the dissipation, modeled by a phenomenological
term based on the second difference of the velocities between adjacent
beads (i.e., a discrete Laplacian) that is raised to a {common} exponent.
This allowed us to augment the standard dynamical model based on Hertzian
forces to encompass
this dissipation effect in (optimal) quantitative agreement with our experiments.  We found that the dissipation prefactor is material-dependent and that the considerably weaker prefactor
of teflon (in comparison to brass and steel) allows one to observe
unambiguously (and for the first time) secondary pulses such as the ones
proposed in Ref.~\cite{LindenbergNesterenkoPRL07}.
Our study also provides a starting point for a potential first-principles
derivation, as well as for future quantitative investigations of this
newly-proposed model.  For example, it would be worth examining the critical
prefactor below which a secondary wave should be expected to emerge, the
interplay of the role of dissipation and plasticity (and a quantitative
incorporation of the latter) in the dynamics, and extensions of the present
considerations to higher-dimensional settings.



\noindent{{\sl \em Acknowledgments.}}
C. D. acknowledges support from {NSF-CMMI 0825345}, and P. G. K. acknowledges
support from NSF-DMS, NSF-CAREER and the AvH Foundation.
{We thank Charles Campbell for useful discussions}.

\vspace{-.2 in}


\end{document}